# Reconfigurable Battery Systems for Enhanced Fast Charging in Electric Vehicles

Jonathan Olivares*, Tyler Depe*, and Rakeshkumar Mahto†
*Department of Electrical and Computer Engineering*
*California State University*
Fullerton, California, USA
*{jonathan_olivares, tylerdepe}@csu.fullerton.edu
†ramahto@fullerton.edu

*Abstract*—The adoption of electric vehicles (EVs) is rapidly growing as a key solution to reducing greenhouse gas emissions. However, prolonged charging times remain a significant barrier to widespread EV usage, especially for individuals without access to fast charging infrastructure. This paper explores the potential of reconfigurable battery systems to reduce EV charging times without compromising battery life. We propose innovative battery pack configurations that dynamically adjust the arrangement of cells to optimize charging performance. Simulations were conducted using MATLAB and Simulink to compare the efficiency of various battery configurations, focusing on charging times, state of charge (SOC), voltage, and current under different conditions. The results demonstrate that connecting more batteries in series through reconfigurability in battery packs can significantly reduce charging times while maintaining operational safety. This study offers insights into how reconfigurable battery designs can provide a practical solution for faster, more efficient home-based EV charging, making EV ownership more accessible and sustainable.

## I. INTRODUCTION

Electric vehicles (EVs) are gaining popularity due to them being considered critical solutions to reduce greenhouse gas emissions. This prompted many local governments to offer incentives for switching from fossil fuel-based vehicles to EVs. Most EVs have lithium-ion batteries because of their higher energy density and long cycle life. However, these batteries have prolonged charging times, discouraging many enthusiastic individuals who might switch to EVs. Therefore, it is essential to develop technologies that can reduce the charging time without affecting the battery life.

DC fast charging has emerged as one of the solutions to fast charging of EV. According to [1], DC fast charging has the capability of charging a vehicle to 80% state-of-charge (SOC) in as little as 10 minutes. However, DC fast charging can negatively impact battery life in the long run due to increased heat generation and accelerated degradation [2], [3]. Liquid and phase change cooling have been developed to prevent heating during charging, which results in battery degradation [4]. Additionally, various algorithms were developed to facilitate fast charging, with improvements focused on enhancing the efficiency and effectiveness of the charging process [5]–[7]. Furthermore, through research, it is shown that compared to the popular constant-current voltage (CC-CV) algorithm for charging [8], boost charging [9] and pulse-based charging [10] are proposed to improve the charging efficiency. Boost charging enables higher current to go to the battery in the initial phase, resulting in fast charging without heating and preventing battery degradation [9]. Similarly, pulse-based charging causes an even distribution of generated heat that prevents formation of hotspot, thus ensuring safer and more efficient charging [10].

Despite these advancements in charging technologies, having them accessible to users at their homes for increased convenience remains too expensive and difficult to implement due to the number of upgrades required. Even having them in a residential complex is too complicated and requires significant infrastructure upgrades. In fact, according to the United States Census Bureau data released on September 2024, around 34.5% of the residents live in apartment complexes [11]. Hence, they do not have access to personal EV charging facilities, making them dependent on public charging.

In this paper, we are exploring the potential of reconfigurable batteries for the EV as a solution to reduce charging time without requiring specialized charging facilities. Our approach aims to optimize battery pack configurations to enhance charging efficiency while maintaining battery longevity. The rest of the paper is organized as follows. Section II outlines this work's objectives, focusing on the investigation and testing of reconfigurable batteries. Section III describes the methodology utilized for the battery pack modeling and simulation. Section IV presents the results of the simulations, followed by a detailed discussion in Section V. Finally, we conclude with future directions for optimizing reconfigurable battery systems in Section VI.

## II. OBJECTIVES

The primary objective of this work is to explore and test the feasibility of a reconfigurable EV battery system to reduce the charging time. This involves evaluating the impact of battery configuration, i.e. the number of batteries connected in series vs parallel, on the charging performance, with a focus on minimizing the charging time while ensuring the longevity of the battery. Additionally, the study seeks to confirm that the proposed reconfigurable system can support sustainable and reliable EV charging solutions, particularly for home-based charging scenarios. Fig. 1 depicts a simple battery pack reconfiguration (2S4P to 4S2P). In this naming convention, (2S4P) refers to 2 battery modules connected in series and 4 sets of these series connected modules connected in parallel. In Fig. 1, one scenario shows two battery modules connected in series, with four of these series-connected modules connected in parallel. In another scenario, as shown in Fig. 1, the configuration is changed so that four battery modules are connected in series, and two of these series-connected modules are connected in parallel.

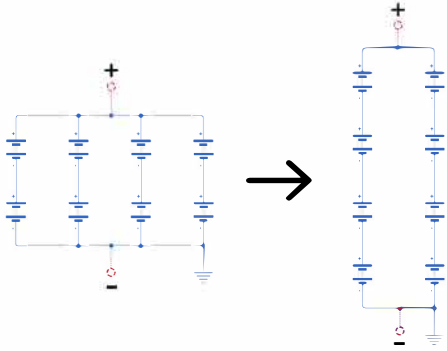

Fig. 1: Simple Battery Pack Reconfiguration (2S4P → 4S2P)

## III. Methodology

This section outlines our approach for modeling the simulation of various battery pack configurations.

### A. Software and Applications

We used MATLAB software to evaluate the performance of various configurations in EV batteries. We especially utilized two robust Matlab applications, including Battery Builder and Simulink [12]. We employed Battery Builder to design specific battery pack configurations and generate Simulink blocks. We operated Simulink to build an electric circuit model to simulate and compare each battery pack configuration.

TABLE I: Tesla Model Y Battery Cell Parameters

| Parameter | Value |
|---|---|
| Geometry | Cylindrical |
| Radius | 0.023 m |
| Height | 0.08 m |
| Mass | 0.355 kg |
| Capacity | 23.35 Ah |
| Energy | 86.5 Wh |

### B. Creating Battery Packs

We complete each assembly step in the following order (cell → parallel assembly → module → module assembly → pack) in Battery Builder to create battery packs. Each battery cells paramenters are shown in Table. I. In the cell section, we defined the geometry and cell properties, such as mass, capacity, and energy. In the parallel assembly section, we defined the number of cells electrically connected in parallel. In the module section, we defined the number of parallel assemblies electrically connected in series. In the module assembly section, we defined the number of modules electrically connected in series. In the pack section, we defined the number of module assemblies connected in series. After we finalize each pack, we generate a Simulink block of the battery pack to allow for simulation on Simulink.

For this work, we created the 2022 Tesla Model Y 92S9P battery pack using specific data sheets and information specifying the pack configuration, as shown in Fig. 2. Then, we reconfigured the battery pack in two ways: a less series more parallel configuration (46S18P) and a more series less parallel configuration (142S5P). As a result, we will experiment with three different battery packs.

### C. Battery Pack Charging/Discharging Circuit Simulation

We use an electric circuit model, Fig. 3, obtained from MATLAB's Simscape Battery Onramp [13], which can simulate the charging/discharging phases, state of charge, voltage, and current for a specific battery pack configuration over a period of time using a particular charging current. The charging and discharging current was altered to imply different charging infrastructure levels. For instance, 15A represents level 1 charging infrastructure, and 30A and 48A represent level 2 charging infrastructure. Below is a description of each Simulink block:

1) Battery Pack (1): generated battery pack using specified parameters from Battery Builder
2) Solver Configuration (1): specifies the solver parameters that the model needs before we can begin simulation
3) Electrical Reference (1): electrical ground
4) Terminator (3): caps blocks whose output ports do not connect to other blocks

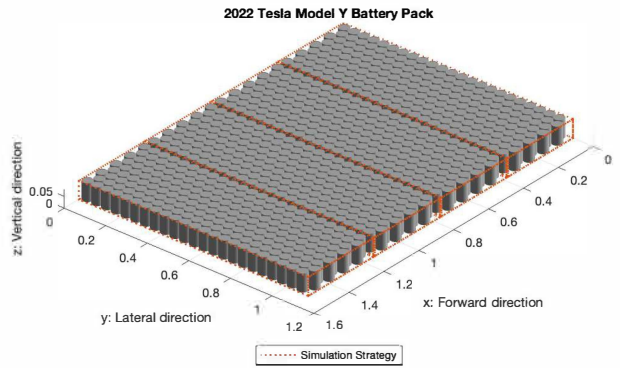

Fig. 2: 2022 Tesla Model Y Battery Pack (92S9P)

5) Controlled Current Source (1): an ideal current source that is powerful enough to maintain the specified current through it regardless of the voltage across the source
6) SOC Estimator (Coloumb Counting) (1): an estimator that calculates the state of charge (SOC) of a battery by using the Coulomb counting method
7) Max of Elements (1): outputs min or max of input
8) Relay (1): toggles between charging and discharging
9) Constant (2): generates a real or complex constant value signal
10) Gain (1): multiplies the input by a constant value (gain)
11) Battery CC-CV (1): implements a constant-current (CC), constant-voltage (CV) charging algorithm for a battery, and for a discharging battery, the block uses the value of the CurrentWhenDischarging input port
12) Unit Delay (1): holds and delays its input by the sample period specified
13) Scope (1): displays signals generated during simulation

For each battery pack (original and modified packs), we simulate the state of charge, voltage, and current using 15 A, 30 A, and 48 A for charging and discharging current for 48 hours. Since this work mainly focuses on the lack of specialized charging at residential homes, we presented our findings for 15A, i.e., level 1 charging, in the results section.

## IV. RESULTS AND DISCUSSION

This section presents the state of charge, voltage, and current performance for each battery pack configuration. The battery packs simulated include the original configuration (92S9P), the less series more parallel reconfiguration (46S18P), and the more series less parallel reconfiguration (142S5P). To represent level 1 charging infrastructure, we showcase the results yielded using 15A as the charging current for 48 hours.

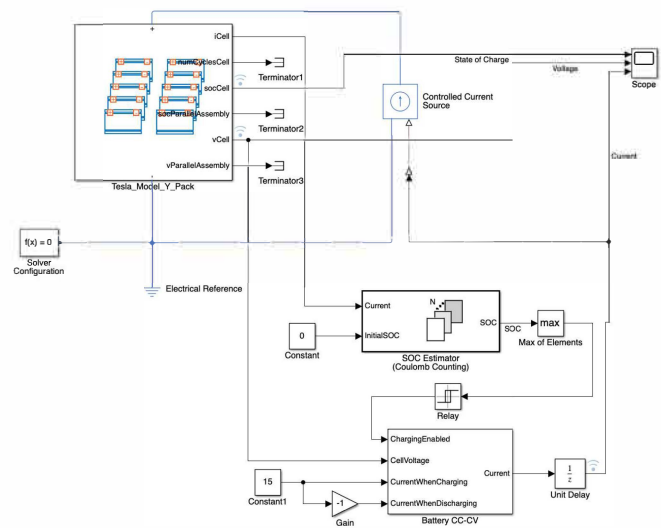

Fig. 3: Battery Pack Charging/Discharging Circuit (Simulink Model)

### A. Charging/Discharging, State of Charge, Voltage, and Current Performance

We consider each battery pack configuration's state of charge, voltage, and current. Fig. 4a) presents the results for the original configuration (92S9P), which shows that it took approximately 16.2 hours to charge from 0% to 100% fully. Fig. 4b) presents the results for the first modified configuration that uses fewer series and more parallel cells (46S18P), showing that it took approximately 32.4 hours to charge from 0% to 100% fully. Fig. 4c) presents the results for the second modified configuration that uses more series and fewer parallel cells (142S5P), demonstrating that it took approximately 9 hours to charge from 0% to 100% fully.

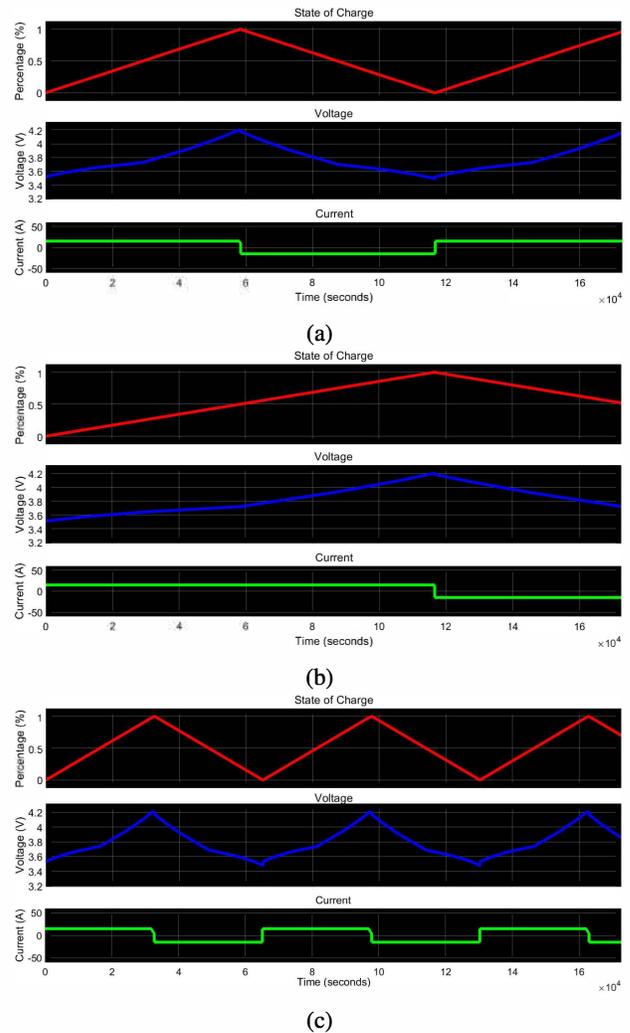

Fig. 4: Comparison of different battery pack configurations (SOC, Voltage, and Current) a) 92S9P Battery Pack (SOC, V, A), b) 46S18P Battery Pack (SOC, V, A), c) 142S5P Battery Pack (SOC, V, A)

### B. Interpretation of Results

To analyze the results of each battery pack configuration, we utilized the Data Inspector tool to perform side-by-side

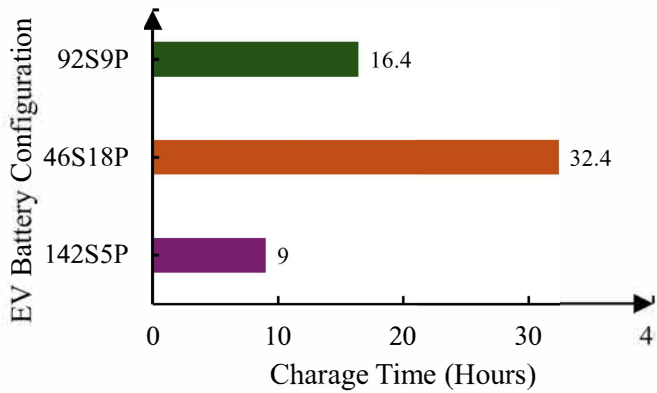

Fig. 5: Comparison between charging time for various configurations

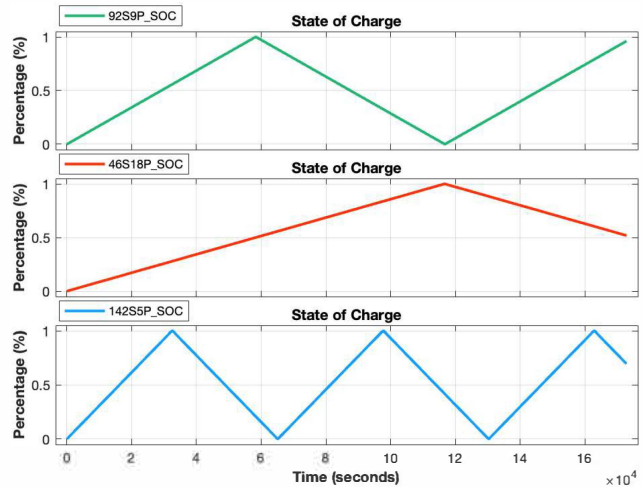

Fig. 6: 2022 Tesla Model Y Battery Pack (92S9P) at 30A comparisons of each state of charge and charging/discharging plot. The fewer series more parallel (46S18P) battery pack reconfiguration took approximately 32.4 hours to charge fully, twice as long as the original configuration, which took 16.2 hours as shown in Fig. 5. Therefore, modifying the original battery pack to this configuration is not efficient for achieving faster charging times.

On the other hand, the more series and fewer parallel cells (142S5P) battery pack reconfiguration took approximately 9 hours to fully charge, which was about 7.2 hours less than the original configuration and much less than the first reconfiguration. As a result, modifying the original battery pack to this configuration is more efficient for faster charging times. Fig. 4 demonstrates a side-by-side comparison of the state of charge for each configuration.

As a side note, we also simulated these battery pack configurations using 30A and 48A. The charging at 30A for Tesla Model Y is shown Fig. 6. The results followed the same pattern of the more series less parallel configuration, achieving faster charging times.

*C. Future Work*

EV battery reconfigurability can provide faster charging speeds, mainly with more series less parallel configurations. As a result, we can use this knowledge for our next steps in this work. We plan to implement a transistor-based mechanism, previously applied to use on solar panels [14], to facilitate battery reconfigurability. The mechanism will allow the EV battery pack to be reconfigured in a specific way to meet a particular objective. For instance, when the EV is charging, the mechanism will reconfigure the battery pack to have more cells electrically connected in series and fewer cells electrically connected in parallel since it will provide faster charging. Then, when the vehicle is in motion or not charging, the mechanism will reconfigure the battery pack back to its original configuration to meet its standard voltage and capacity operating requirements. However, to facilitate reconfiguration of the EV battery, a newer type of transistors made from advanced semiconductor materials such as silicon carbide (SiC) or gallium nitride (GaN), known for their higher current-carrying capacity and switching efficiency, will be identified and explored.

Additionally, a machine learning-based algorithm will be developed to reconfigure the battery during charging dynamically, optimizing the process to significantly reduce charging time and enhance efficiency. As we progress, we will consider more factors like temperature in the research process to better understand all the variables that come with EV batteries that manufacturers also consider. That way, we can include it as an input feature in the machine learning model to facilitate a fast charging.

## V. CONCLUSION

This work explores next-gen reconfigurable EV batteries to enhance at-home EV charging. This includes designing various battery pack configurations and using an electric circuit model to simulate and compare the performance of each battery pack configuration, specifically the charging/discharging phases, state of charge, voltage, and current. We utilize Battery Builder and Simulink in MATLAB to accomplish these tasks.

Our findings indicate that reconfiguring a battery pack to have more cells electrically connected in series and less cells electrically connected in parallel can lead to enhanced charging efficiency. As a result, reconfigurable battery designs can make EVs more practical for households with limited charging infrastructure, allowing for improved home charging feasibility. Additionally, efficient battery configurations can contribute to the goal of transitioning to a greener future by making EVs more practical and sustainable for everyday use, which advances sustainability and accessibility.

For our future work, we plan to employ a transistor-based mechanism to facilitate battery reconfigurability and optimize a machine-learning model to yield the best results for the application of EV batteries. This mechanism will allow faster at-home EV charging and regular operation without compromising performance and range. Additionally, we will consider disregarded factors, such as temperature effects, aging, leaking, etc., to understand all variables associated with EV batteries further.

### Acknowledgment

The authors are grateful to the Undergraduate Research Opportunity Center (UROC) at California State University, Fullerton (CSUF), for the monetary and professional development support to the undergraduate students involved in this work.